\documentstyle[12pt]{article}
\begin{document}
\def\thefootnote{\fnsymbol{footnote}}
\setcounter{footnote}{1}
\newcommand{\be}{\begin{equation}}
\newcommand{\ee}{\end{equation}}
\newcommand{\bdm}{\begin{displaymath}}
\newcommand{\edm}{\end{displaymath}}
\newcommand{\bea}{\begin{eqnarray}}
\newcommand{\eea}{\end{eqnarray}}
\newcommand{\bda}{\begin{\displaymath}\begin{array}{rl}}
\newcommand{\eda}{\end{array}\end{displaymath}}
\newcommand{\co}{\; \; ,}
\newcommand{\fs}{\; \; .}
\newcommand{\no}{\nonumber \\}
\newcommand{\al}{&\!\!\!\!}
\newcommand{\cl}{\cal L}
\newcommand{\integral}{\int\!d^4\!x}
\newcommand{\eps}{\epsilon^{\mu \nu \alpha \beta}}
\newcommand{\nab}{\bf \nabla \rm}
\newcommand{\th}{\theta}
\newcommand{\vr}{\rightarrow}
\newcommand{\Tr}{\rm T}
\newcommand{\Lo}{\rm L}
\newcommand{\la}{\it a}
\newcommand{\hs}{\hspace{0.25em},\hspace{0.25em}}
\newcommand{\hsl}{\hspace{0.1em},\hspace{0.1em}}
\newcommand{\WZW}{\mbox{\scriptsize W\hspace{-0.1em}ZW}}
\newcommand{\EM}{\mbox{\scriptsize E\hspace{-0.1em}M}}
%
%
\begin{titlepage}
\begin{raggedleft}
THES-TP 95/03\\
April.1995\\
\end{raggedleft}
\vspace{2em}
\begin{center}
{\Large\bf{ Application of the
Faddeev-Jackiw\\
\vspace{0.25em}
formalism to the gauged WZW model}}\\
\vspace{2em}
{\large J.E.Paschalis and P.I.Porfyriadis}
\footnote{On leave from the Tbilisi State University, Tbilisi,
Georgia.}
\vspace{1em}\\
Department of Theoretical Physics, University of
Thessaloniki,\\Gr-54006 \hs Thessaloniki \hs Greece\\
\vspace{1em}
{\scriptsize PASCHALIS@OLYMP.CCF.AUTH.GR\\
PORFYRIADIS@OLYMP.CCF.AUTH.GR}
\end{center}
\vspace{2em}
\begin{abstract}
The two-flavour Wess-Zumino model coupled to electromagnetism
is treated as a constraint system using the Faddeev-Jackiw
method. Expanding into series of powers of the pion fields and
keeping terms up to second and third order we obtain
Coulomb-gauge Lagrangeans containing non-local terms.
\end{abstract}
\end{titlepage}
%
%
\section{Introduction}
\label{intro}
The Faddeev-Jackiw method
\cite{F-J} avoids the separation of the
constraints into first and second class and gives us a simple
and straightforward way to deal with constraint systems. A brief
outline of this technique is given below. Let $H(p,q)$ be
the Hamiltonian describing the dynamics of a certain system.
We can always construct a first order in time derivatives
Lagrangean whose configuration space coincides with the
Hamiltonian  phase space. This can be done by enlarging
the $n$ dimensional configuration space to a $2n$ dimensional
configuration space. We define the new coordinates $\xi^i$ ,
as follows
\bea
  \xi^i \al=\al p_i\hs\hspace{2em} i=1,...,n \nonumber \\
  \xi^i \al=\al q^i\hs\hspace{2em} i=n+1,...,2n \nonumber
\eea
Then the Lagrangean of the system can be written as
\be
  {\cl}(\xi,\dot{\xi})=\frac{1}{2}\xi^i\omega_{ij}\dot{\xi^j}
                   - H(\xi)\co
\ee
after dropping a total time derivative. The matrix
$ \omega_{ij} $  is given by
\bdm
 \omega_{ij}=\left( \matrix{0 & I \cr -I& 0 \cr} \right)_{ij}
\edm
and has an inverse $\omega^{ij}$. The system of the example has
no constraint and the evolution of $\xi$ is given as follows
using the Euler-Lagrange equations
\be
  \dot{\xi}^i = \omega^{ij}\frac{\partial}{\partial\xi^j}
                           H(\xi)\fs
\ee

For a general unconstrained system described by the Lagrangean
\be
   {\cl}(\xi,\dot{\xi}) = a_i(\xi)\dot{\xi^i} -
                      H(\xi)\hs\hspace{2em}i=1,...,N
\ee
with arbitrary $a_i(\xi)$ , the Euler-Lagrange equations
are given by
\be
   f_{ij}\dot{\xi^j} = \frac{\partial}{\partial \xi_i}H\co
\ee
where $f_{ij}=\frac{\partial}{\partial \xi_i}a_j -
\frac{\partial}{\partial \xi_j}a_i$ is an invertible $2n\times
2n$  matrix. Then according to Darboux's theorem we can construct
a coordinate transformation $Q^i(\xi)$ so that the canonical
one-form $a_{i}(\xi)d\xi^i$ defined in the first part of (3)
acquires the "diagonal" form given in (1)
\be
  a_i(\xi)d\xi^i = \frac{1}{2}Q^k(\xi)\omega_{kl}dQ^{l}(\xi)\fs
\ee

In the case that the Lagrangean (3) describes a constraint
system the matrix $f_{ij}$ is singular.
The Darboux's theorem can still be applied for the maximal
$2n\times 2n$ nonsingular subblock of $f_{ij}$. The
Lagrangean (3) transforms as follows
\be
  {\cl}(Q,\dot{Q},z)=\frac{1}{2}Q^k\omega_{kl}\dot{Q^l} -
                 H'(Q,z)\hs \hspace{2em} k,l=1,...,2n
\ee
where  $z$ denote the $N-2n$ coordinates that are left unchanged.
Some of the $z_i$ may appear non-linearly and some linearly
in $H'(Q,z)$. Then using the Euler-Lagrange equation for these
coordinates we can solve for as many $z_i$'s
as possible in term of $Q^i$'s and other $z_i$'s and replace
back in $H'(Q,z)$ so finally we are left only with linearly
occuring $z_i$'s. After this step is completed the Lagrangean
(6) can be written as
\be
  {\cl}=\frac{1}{2}Q^k\omega_{kl}\dot{Q^l}-V(Q)-z_i\Phi^i(Q)\co
\ee
where we see that $z_{i}$ become the Lagrange multipliers and
$\Phi^{i}(Q)$ are the constraints. By solving the equations of
constraints and replacing back in   (7) we reduce the number of
Q's and we end up with a Lagrangean which has the structure
given in (3). Then the whole procedure can be repeated again
until all constraints are eliminated and we are left
with an unconstrained Lagrangean whose canonical one form is
diagonal. Now the canonical quantization rules can be applied to
the new canonical coordinates, which we rename $p_i\hs q^i\hs
i=1,...,n$.

The FJ approach to constraint systems has by now been
successfully used in various fields
\cite{Govaerts,Horta,Kulsh,Montani,Barcelos,Jun,Jacob,
Cronstrom,Wotzasek} . Our intention is
to apply it to the two-flavour WZW model coupled to
electromagnetism. This model describes
very well the low energy interactions among pions and pion with
photons including those related to the axial anomaly.
%
%
\section{Expanding up to second order in the pion fields}
\label{exp-sec}

The effective action of the model
\cite{Witten,Pak,Donoghue} given by
\bea
  \Gamma_{eff}(U,A_\mu)\al =\al {\Gamma}_{\EM} (A_\mu) +
                       \Gamma_\sigma (U,A_\mu) +
                       {\Gamma}_{\WZW} (U,A_\mu)\co\\
  {\Gamma}_{\EM} (A_\mu)\al=\al -\frac{1}{4}\integral
                 F_{\mu \nu} F^{\mu \nu}\co\nonumber\\
  \Gamma_\sigma (U,A_\mu)\al =\al -\frac{f_\pi^2}{16}
                 \integral \mbox{tr}\,(R_\mu R^\mu)\no
                 \al=\al-\frac{f_\pi^{2}}{16}
                 \integral \mbox{tr}\,(r_\mu r^\mu) +
                 \frac{if_\pi^{2}e}{8}
            \integral A_\mu \mbox{tr}\,[Q(r_\mu-l_\mu)]
                 \no \al+\al \frac{f_\pi^{2}e^2}{8}
                 \integral A_\mu A^\mu \mbox{tr}\,
                 (Q^2-U^\dagger QUQ)\co \nonumber\\
  {\Gamma}_{\WZW} (U,A_\mu)\al =\al -\frac{N_{c}e}{48\pi^2}
                 \integral \eps A_\mu \mbox{tr}\,
                  [Q(r_\nu r_\alpha r_\beta +
                  l_\nu l_\alpha l_\beta)]\no
                  \al+\al \frac{iN_{c}e^2}{24\pi^2}
                  \integral \eps A_\mu(\partial _\nu A_\alpha)
                  \mbox{tr}\,[Q^2(r_\beta+l_\beta) \no
                  \al+\al \frac{1}{2}QU^\dagger QUr_\beta
                 +\frac{1}{2}QUQU^\dagger l_\beta] \nonumber\co
\eea
where
\bea
   U\al=\al\exp\,(2i\th_a\tau_a/f_\pi)\co\hspace{1em}
   r_\mu=U^\dagger\partial_\mu U\co\hspace{1em}
   R_\mu=U^\dagger D_\mu U\co\nonumber \\
   l_\mu\al=\al(\partial_\mu U)U^\dagger \co\hspace{1em}
   L_\mu=(D_\mu U)U^\dagger\fs\nonumber
\eea
See Appendix for notation.
In the three flavor WZW model the term
\bdm
   {\Gamma}_{WZW}(U)=-\frac{iN_c}{240\pi^2}
                \int\!d^5\!x\epsilon^{ijklm}
                \mbox{tr}\,(l_{i}l_{j}l_{k}l_{l}l_{m})\co
\edm
should also be included but this term vanishes in our case. The
effective action (8) is by construction gauge invariant under
$U_{EM}(1)$ gauge transformation. Applying the FJ method to the
full case seems quite difficult so we use expansion into powers of
the pions fields $\th_a$
\be
   U=1+\frac{2i}{f_\pi}\theta_{a}\tau_{a}+...
\ee
We substitute into (8) and
keep terms up to $2^{nd}$ order. After calculating the traces
the effective Lagrangean of the two-flavor Wess-Zumino model
coupled to electromagnetism is written as follows
\bea
  {\cl}_{eff}\al=\al{\cl}_{EM}+
                 {\cl}_{\sigma}^{(2)}+{\cl}_{WZW}^{(2)}
                 +O(\theta^3)\co\\
  {\cl}_{EM}\al=\al-\frac{1}{4}
                F_{\mu \nu}F^{\mu \nu}\co\nonumber\\
  {\cl}_{\sigma}^{(2)}\al=\al \frac{1}{2}
               \partial_\mu \theta_{a}\partial^\mu\th_{a}+
               eA^\mu(\th_{2}\partial_\mu\theta_{1}-
                      \theta_{1}\partial_\mu\theta_{2})
               +\frac{e^2}{2}A_\mu A^\mu
                (\th_{1}^{2}+\th_{2}^{2})\co\nonumber\\
  {\cl}_{\WZW}^{(2)}\al=\al-\frac{N_{c}e^2}{12\pi^{2}f_\pi}
               {\eps}A_\mu(\partial_\nu A_\alpha)\partial_\beta
               \theta_{3}\fs\nonumber
\eea
Then in the non-covariant notation
\bea
   {\cl}_{eff}\al=\al-\bf E \cdot \dot{\bf A}\rm+
               \frac{1}{2}\dot{\theta_{\la}}^{2}+
               eA_{0}(\theta_{2}\dot{\theta_{1}} -
                      \theta_{1}\dot{\theta_{2}})+
               A_{0}\bf \nabla \cdot E\rm\no
       \al+\al \frac{e^2}{2}(A_{0}^{2}-{\bf A}^{2}\rm)
               (\th_{1}^{2}+ \theta_{2}^{2})-
               \frac{1}{2}(\bf E^{2}+ B^{2} +
               (\nabla\rm\th_{\la})^{2})\no\al+\al
               e\bf A\cdot\rm(\th_{2}\bf\nabla
                \rm\th_{1}-\th_{1} \bf \nabla\rm \th_{2})+
                \frac{N_{c}e^2}{6\pi^{2}f_\pi}
                (\bf E \cdot B)\rm \theta_{3}\co
\eea
where $\bf E=-\dot{\bf A}-{\nabla}\rm A_{0},\hspace{1em}
\bf B=\bf{\nabla \times A}\fs$

The canonical momenta conjugate to $\theta_{a}$ are given by
\bdm\begin{array}{lllll}
\vspace{0.5em}
  p_{1}=\frac{\partial {\cl}_{eff}}{\partial \dot{\theta}_1}=
        \dot{\theta}_{1}+eA_{0}\theta_{2}\;\;,\\
\vspace{0.5em}
  p_{2}=\frac{\partial {\cl}_{eff}}{\partial \dot{\theta}_2}=
         \dot{\theta}_{2}-eA_{0}\theta_{1}\;\;,\\
\vspace{0.5em}
  p_{3}=\frac{\partial {\cl}_{eff}}{\partial \dot{\theta}_3}=
        \dot{\theta}_{3}\fs\end{array}
\edm
In the enlarged configuration space with coordinates $-E_a\hs
A_a\hs p_a\hs\theta_a\hs$ $\hspace{0.5em} a=1,2,3 $ the
effective Lagrangean (11) can be written after dropping a
total time derivative as an expression
first order in time derivatives of the coordinates, having
the structure given in (7)
\bea
  {\cl}_{eff}\al=\al-\bf E\cdot\dot{A}\rm+
                 p_{\la}\dot{\theta_{\la}}-
          H^{(2)}(\bf E,A,\rm p_{\la},\theta_{\la})
                -A_{0}(\rho^{(2)}-\bf\nabla\cdot E\rm)
                +O(\th^3)\co\\
\vspace{1em}
          H^{(2)} \al=\al\frac{1}{2}[\bf E^2+B^2
              +(\nabla \rm\th_{\la})^{2}+p_{\la}^{2}]
              +e\bf A\cdot\rm(\th_{1}\nab \th_{2}-
                              \theta_2 \nab \theta_1)+
               \frac{e^2}{2}\bf A^2\rm
                  (\theta_{1}^{2}+\theta_{2}^{2})\no
         \al-\al\frac{N_{c}e^2}{6\pi^2 f_\pi}\bf
               (E\cdot B)\rm\th_{3}\co\nonumber\\
\vspace{1em}
  \rho^{(2)}\al=\al e(p_{2}\theta_1-p_{1}\theta_{2})
                 \fs\nonumber
\eea
We see that  the  scalar  potential $A_{0}$ is the Lagrange
multiplier  and $\rho^{(2)}-\bf \nabla\cdot E\hspace{0.25em}$
is the constraint. In order to solve the equation of the
constraint
\be
   \nab \cdot \bf E\rm-\rho^{(2)}=0\co
\ee
we decompose the electric field $\bf E$
and the vector potential $\bf A$ into transverse and
longitudinal components
\bdm
  \bf E=E_{\Tr}+E_{\Lo}\hs \hspace{1em} A=A_{\Tr}+A_{\Lo}\hs
\edm
\bdm
   \bf \nabla \cdot E_{\Tr}=\rm 0 \hs
   \bf \nabla \times E_{\Lo}=\rm 0 \hs
   \bf \nabla \cdot A_{\Tr}=\rm 0 \hs
   \bf \nabla \times A_{\Lo}=\rm 0\fs
\edm
Then (13) implies that
\be
   \bf E_{\Lo}\rm=\frac{\nab}{\nab^2}\rho^{(2)}\fs
\ee
Substituting into (12) we obtain apart from total spatial
derivatives
\bea
  {\cl}_{eff} \al=\al -\bf E_{\Tr}\cdot\dot{A}_{\Tr}\rm
       +\rho^{(2)}\frac{\nab}{\nab^2}\cdot\dot{\bf A}_{\Lo}\rm
       +p_{\la}\dot{\th}_{\la}\no\al-\al \frac{1}{2}
        [\bf E_{\Tr}^2+ B^2-\rm \rho^{(2)}\frac{1}{\nab^2}
         \rho^{(2)}+(\nabla \rm\theta_{\la})^{2}+p_{\la}^{2}]
         \no \al-\al e\bf A\cdot\rm
         (\th_{1} \nab \th_{2}-\th_2 \nab \th_1)
         - \frac{e^2}{2}\bf A\rm^2 (\th_{1}^{2}+\th_{2}^{2})+
         \frac{N_{c}e^2}{6\pi^2 f_\pi}
         (\bf E\cdot B\rm)\th_3\fs
\eea
We see that the longitudinal part of the
vector potential $\bf A_{\Lo}$ , enters the canonical one-form
of the Lagrangean in an uncanonical way.
In order to diagonalize the canonical one-form in (15) we perform
the following Darboux's transformations
\bdm
   p_1\vr p_1 \cos{\alpha}+p_2\sin{\alpha} \hs \hspace{1em}
  \theta_1 \vr \theta_1 \cos{\alpha}+\theta_2 \sin{\alpha}\hs
\edm
\be
   p_2 \vr p_2 \cos{\alpha}-p_1\sin{\alpha}\hs \hspace{1em}
  \theta_2 \vr \theta_2 \cos{\alpha}-\theta_1 \sin{\alpha}\hs
\ee
\bdm
   p_3 \vr p_3 \hs \theta_3 \vr \theta_3 \hs
  \bf E_{\Tr} \vr E_{\Tr}\hs A_{\Tr} \vr A_{\Tr}\hs
\edm
where $\alpha = e\frac{\nab}{\nab^2}\cdot\bf A_{\Lo}$
This choice of coordinate transformations leads to the
cancelation of $\bf A_{\Lo}$ and fixes the gauge of the
Lagrangean to the Coulomb-gauge
\bea
  {\cl}_{eff}\al=\al-\bf E_{\Tr}\cdot\dot{A}_{\Tr}+
             \rm p_{\la}\dot{\theta_{\la}}-
   H_{C}^{(2)}
     (\bf E_{\Tr},A_{\Tr},\rm p_{\la},\theta_{\la})\co\\
   H_{C}^{(2)}
     (\bf E_{\Tr},A_{\Tr},\rm p_{\la},\theta_{\la})
          \al=\al\frac{1}{2}[\bf E_{\Tr}^2+B^2\rm-
                  \rho^{(2)}\frac{1}{\nab^2}
                  \rho^{(2)}+(\nab \rm\theta_{\la})^{2}
                  +p_{\la}^{2}]\no
          \al+\al e\bf A_{\Tr}\cdot\rm
                 (\th_{1} \nab \th_{2}-\th_2 \nab \th_1)
             + \frac{e^2}{2}\bf A_{\Tr}^2\rm
                    (\th_{1}^{2}+\th_{2}^{2})
        \no\al-\al\frac{N_{c}e^2}{6\pi^2 f_\pi}
                    [\bf E_{\Tr}\cdot B\rm +
       (\frac{\nab}{\nab^2}\rm \rho^{(2)})\bf\cdot B\rm]\th_3
                    \nonumber\fs
\eea
We see that only the physical transverse components of the
vector potential enter in the expression of the Lagrangean.
( Note that $ \bf B=\nabla\times A_{\Tr} $ ).
%
%
\section{Keeping third order terms}
\label{exp-third}

We now proceed with the expansion and keep terms up to third
order in the Goldstone boson fields
\be
    {\cl}_{eff}={\cl}_{EM}+{\cl}_{\sigma}^{(2)}+
                {\cl}_{WZW}^{(2)}+{\cl}_{WZW}^{(3)}
                +O(\theta^4)\co
\ee
where the expressions for ${\cl}_{\sigma}^{(2)}$ and
${\cl}_{WZW}^{(2)}$ are given in (10) and
\bea
   {\cl}_{WZW}^{(3)}\al=\al-\frac{N_{c}e}{3\pi^{2}f_{\pi}^3}
               {\eps}(\partial_\mu A_\nu)
               (\theta_1\partial_\alpha \theta_2 -
                \theta_2 \partial_\alpha \theta_1)
               \partial_\beta \theta_3\no
        \al+\al \frac{2N_{c}e^2}{9\pi^{2}f_{\pi}^3}
                {\eps}(\partial_\mu A_\nu)
                      (\partial_\alpha A_\beta)
                (\theta_{1}^2+\theta_{2}^2)\theta_3\no
        \al-\al \frac{N_{c}e^2}{3\pi^{2}f_{\pi}^3}
                {\eps}  A_\mu (\partial_\nu A_\alpha)\theta_3
                \partial_\beta (\theta_{1}^2+\theta_{2}^2)
                \nonumber\fs
\eea
We see that the Wess-Zumino-Witten
term is the only term that contributes to this order.
Next we derive the expressions for the canonical momenta $p_{a}$
conjugate to the pion fields $\th_{a}$ from the new
Lagrangean (18). Then after long mathematical manipulations
and dropping total derivatives we obtain the following
expression for ${\cl}_{eff}$ in the enlarged configuration space
\vspace{1em}
\be
   {\cl}_{eff}=-\bf E\cdot \dot{A}\rm +
                p_{\la}\dot{\theta_{\la}}
               - H^{(2)}-H^{(3)}-A_{0}(\rho^{(2)}+\rho^{(3)}
               -\bf\nabla\cdot E\rm)+O(\theta^4)\co
\vspace{1em}
\ee
\vspace{1em}
where the expression for $H^{(2)}$ is given in (12) and
\bea
\vspace{1em}
   H^{(3)}\al=\al-\frac{N_{c}e}{3\pi^{2}f_{\pi}^3}
              (\bf E\times\nab \theta_3-p_3 \bf B\rm)
              \cdot(\th_1\nab \th_2-\th_2\nab\th_1)\no
      \al+\al \frac{4N_{c}e^2}{9\pi^{2}f_{\pi}^3}
              (\bf E\cdot B\rm)
              ( \th_{1}^2+\th_{2}^2)\th_3 -
              \frac{N_{c}e^2}{3\pi^{2}f_{\pi}^3}
              (\bf E\times A\rm)
              \cdot [\nab(\th_{1}^2+\th_{2}^2)]\th_3\no
      \al-\al \frac{2N_{c}e^2}{3\pi^{2}f_{\pi}^3}
              (\bf A\cdot B\rm)(p_1\th_1+p_2\th_2)\th_3
              - \frac{N_{c}e}{3\pi^{2}f_{\pi}^3}
              (\bf B\cdot\nab\th_3)
              (p_2\th_1-p_1\th_2)\co\nonumber\\
  \rho^{(2)}\al=\al e(p_2\th_1-p_1\th_2)\co\nonumber\\
  \rho^{(3)}\al=\al-\frac{N_{c}e^2}{3\pi^{2}f_{\pi}^3}
                \nab\cdot[\bf B\rm (\th_{1}^2+\th_{2}^2)\th_3]
                \co\nonumber
\eea
$A_{0}$ is again the Lagrange multiplier and the equation of
the constraint is given by
\be
   \bf \nabla\cdot E\rm-(\rho^{(2)}+\rho^{(3)})=0\fs
\ee
This equation has similar structure as in the previous case (13)
with only an extra term of the third order in the pion fields in
the divergence. So we proceed similarly and decompose $\bf E$ and
$\bf A$ into transverse and longitudinal components.Then (20)
implies that
\bdm
   \bf E_{\Lo}=\rm \frac{\nab}{\nab^2}(\rho^{(2)}+\rho^{(3)})\fs
\edm
Substituting into (19) we end up with a Lagrangean first order in
time derivatives of the coordinates whose canonical one-form
\bdm
  -\bf E_{\Tr}\cdot \dot{A}_{\Tr}\rm
  +\rho^{(2)}\frac{\nab}{\nab^2}\cdot\dot{\bf A}_{\Lo}\rm
  +\rho^{(3)}\frac{\nab}{\nab^2}\cdot\dot{\bf A}_{\Lo}\rm
  +p_{\la} \dot{\th}_{\la}\co
\edm
must be diagonalized. To do this we proceed in two steps.
First we perform the Darboux's transformations given in (16)
as in the previous case.
This transformations lead to partial diagonalization.
We obtain a new Lagrangean whose canonical one-form is given by
\bdm
  -\bf E_{\Tr}\cdot \dot{A}_{\Tr}\rm
  +\rho^{(3)}\frac{\nab}{\nab^2}\cdot\dot{\bf A}_{\Lo}\rm
  +p_{\la} \dot{\th}_{\la}\co
\edm
In order to proceed with the second step we must write the term
$\rho^{(3)}\frac{\nab}{\nab^2}\cdot\dot{\bf A}_{\Lo}\rm$ in a
more suitable way
\bdm
  \rho^{(3)}\frac{\nab}{\nab^2}\cdot\dot{\bf A}_{\Lo}\rm =
   tot.der. - \frac{N_{c}e^2}{3\pi^2 f_{\pi}^3}
   [(\nab{\phi}\times\bf A_{\Lo})\cdot\dot{A}_{\Tr}\rm +
    (\bf B\cdot A_{\Lo}\rm)\dot{\phi}]\co
\edm
where $\phi=(\th_{1}^2+\th_{2}^2)\th_3\fs$

Now we perform the second transformation
\bdm
 \begin{array}{l}
\vspace{0.5em}
  \bf E_{\Tr}\vr E_{\Tr}-\rm \frac{N_{c}e^2}{3\pi^{2}f_{\pi}^3}
                        \nab[(\th_{1}^2+\th_{2}^2)\th_3]\bf
                        \times A_{\Lo}\;\;,\\
\vspace{0.5em}
  p_1 \vr p_1 +\frac{2N_{c}e^2}{3\pi^{2}f_{\pi}^3}
               (\bf B\cdot A_{\Lo}\rm) \th_1 \th_3\;\;,\\
\vspace{0.5em}
  p_2 \vr p_2 +\frac{2N_{c}e^2}{3\pi^{2}f_{\pi}^3}
               (\bf B\cdot A_{\Lo}\rm) \th_2 \th_3\;\;,\\
\vspace{0.5em}
  p_3 \vr p_3 +\frac{N_{c}e^2}{3\pi^{2}f_{\pi}^3}
               (\bf B\cdot A_{\Lo}\rm)(\th_{1}^2+
                 \th_{2}^2)\;\;,\\
\vspace{0.5em}
     \bf A_{\Tr} \vr A_{\Tr}\rm\hs
     \th_1\vr\th_1\hs
     \th_2\vr\th_2\hs
     \th_3\vr\th_3\fs
 \end{array}
\edm
This transformation completes the diagonalization of the
canonical one-form and we end up with a Lagrangean
where the longitudinal part of
the vector potential $A_{\Lo}$ cancels out exactly
\be
  {\cl}=-\bf E_{\Tr}\cdot \dot{A}_{\Tr}\rm +
         p_{\la}\dot{\theta_{\la}}
       - H_{C}^{(2)}-H_{C}^{(3)}+O(\theta^4)\co
\ee
where
\bea
   H_{C}^{(3)}\al=\al-\frac{N_{c}e}{3\pi^{2}f_{\pi}^3}
              (\bf E_{\Tr}\times \nab \theta_3-p_3 \bf B\rm)
              \cdot(\th_1 \nab \th_2-\th_2\nab\th_1)\no
      \al+\al \frac{4N_{c}e^2}{9\pi^{2}f_{\pi}^3}
              (\bf E_{\Tr}\cdot B\rm)
              (\th_{1}^2+\th_{2}^2)\th_3 -
              \frac{N_{c}e^2}{3\pi^{2}f_{\pi}^3}
              (\bf E_{\Tr}\times A_{\Tr}\rm)\cdot
              [\nab(\th_{1}^2+\th_{2}^2)]\th_3\no
      \al-\al \frac{2N_{c}e^2}{3\pi^{2}f_{\pi}^3}
              (\bf A_{\Tr}\cdot B\rm)
              (p_1\th_1+p_2\th_2)\th_3 -
              \frac{N_{c}e}{3\pi^{2}f_{\pi}^3}
              (\bf B\cdot\nab\th_3)
              (p_2\th_1-p_1\th_2)\co\nonumber
\eea
and the expression for $H_{C}^{(2)}$ is the given in (17).

We see that application of the Faddeev-Jackiw method for
constraint systems to the two flavour Wess-Zumino-Witten model
coupled to electromagnetism leads up to second and
third order in the pion
fields to Coulomb-gauge Lagrangeans including nonlocal
interaction terms as given in (17). Futher application of the
method to the three flavour case is currently under
investigation.
%
%
\section{Appendix}
\label{app}

Our metric is $g_{\mu \nu}=diag(1,-1,-1,-1)\hs Q=diag(2/3,-1/3)$
is the charge matrix,
 $D_\mu=\partial_\mu + ieA_\mu [Q,\hspace{0.4em}]$
denote the covariant derivative.
By $\tau_{a}\hs a=1,2,3$  we denote
Pauli matrices. We choose $ e>0 $ so that the electric charge of
the electron is $-e$. We define $ \epsilon^{0123}=1$ .

\end{document}